\documentclass[11pt]{article} 

\usepackage{newcent} 
\usepackage{hyperref}
\usepackage{amsmath}
\usepackage{pdflscape}
\usepackage{arydshln}
\usepackage{epstopdf}
\usepackage{booktabs}
\usepackage{authblk}
\usepackage{multirow}
\usepackage{graphicx}

\author[1]{Bismark~Singh}
\author[2]{David~Pozo}
\affil[1]{Discrete Mathematics \& Optimization, Sandia National Laboratories, Albuquerque, NM 87185, USA}
\affil[2]{Center for Energy Systems, Skolkovo Institute of 
Science and Technology, Moscow, Russia}

\title{A Guide to Solar Power Forecasting using ARMA Models}

\begin{document}
%

%
%
%

\maketitle

\section*{Abstract}
	We describe a simple and succinct methodology to develop hourly auto-regressive 
	moving average (ARMA) models to forecast power output from a photovoltaic solar generator. 
	We illustrate how to build an ARMA model, to use  statistical tests to validate it, and construct hourly samples. 
	The resulting model inherits nice properties for embedding it into more sophisticated operation and planning models, while at the same time showing relatively good accuracy. Additionally, it represents a good forecasting tool for sample generation for stochastic energy optimization models.


\section{Introduction}
Increasing penetration of renewable energy sources, such as wind and solar, in 
the electricity grid requires an accurate representation of the 
uncertainty to guarantee feasibility of the system operations, and 
for efficiently planning new transmission lines and generation capacities. To 
address these challenges, advanced decision-making models that lie at the
interface of statistics and operation research have been widely 
explored.  Representation of the  uncertainty in renewable energy is typically  
done by either using \textit{samples} or a  \textit{set representation} from the 
underlying stochastic process. The former generally requires forecasting tools 
for generating synthetic samples or scenarios that are used for feeding 
decision-making optimization models~\cite{kleywegt2002sample}. The latter 
requires a simple representation of the stochastic process in 
order to embed it into more sophisticated decision-making 
tools~\cite{lorca2015adaptive}. In both cases, but especially so in the latter, 
complex forecasting models result in models that are hard to integrate. 

In this article, we present a forecasting model that is easy-to-embed into more 
sophisticated decision-making models, which at the same time also serves as a 
tool for generating samples of renewable energy forecasts.  In 
particular, we focus on 
forecasting  hourly photovoltaic  (PV) solar power generation, but the 
methodology is not 
limited to this technology.  Solar power differs from wind power due 
to its diurnal nature, and can have much greater  ramps than 
wind~\cite{graabak2016variability}. 

Forecasting methods for solar power are broadly divided 
into two categories: (i) physics-based models---these models predict solar 
power 
from numerical weather predictions and solar irradiation data, and (ii) 
statistical models---these 
models forecast 
solar power directly from historical data. Comparisons of these two 
methods are available; see, 
e.g.,~\cite{huang2010comparative}. There are other 
approaches available as well which combine these two 
methods~\cite{dong2015novel}. 
In this article, we center on  statistical  methods alone, and specifically the 
use of auto-regressive moving average (ARMA) 
models to develop our forecasts. 

Despite 
their limitations~\cite{yang2018history}, ARMA models
are widely used to forecast wind power~\cite{duran2007short}, as well as solar 
power~\cite{mora1998multiplicative}, because of their ease of 
implementation and parameter selection.  Yet, accurate and 
fast methods to generate solar power scenarios are often unavailable or 
significantly complex, and normal approximations are frequently used; see, 
e.g.,~\cite{su2014stochastic}. 
Here, 
we describe a summary of the methodology to forecast solar power using 
ARMA models. The software codes and generated scenarios are 
available on request. The presented models can 
be applied either to a local PV generating plant or  at the regional 
level.

The main contribution of this article is to provide a step-by-step approach and 
easy-to-implement ARMA model to forecast PV solar power generation. The 
proposed model is able to capture the important statistical features of the 
parameters, while maintaining simplicity. The model allows modelers to embed
it into more complex decision-making structures, statisticians to have an 
all-in-one place ARMA model design for PV power generation, and policy makers and electrical engineers to have a scenario generation tool. 

\section{Methodology} \label{sec:methodology}

We take hourly year-long historical solar power output 
from a site described in~\cite{golestaneh2016generation}. 
This zone has an altitude of 595m, a nominal power of 1560~MW, a panel 
tilt 
of $36^\circ$, and a $38^\circ$ clockwise panel orientation from the north. 
Further installation specifics are available in zone~1 from Table~1 
of~\cite{golestaneh2016generation}, while technical specifications are 
available 
in~\cite{technical}.  We use approximately nine months of data 
for training. The data does not have any solar power for the ten hours 
[20:00-5:00], and hence we restrict the forecasts in these hours to be zero as 
well. Equivalently, a criteria based on the solar zenith angle can be used; 
i.e., $0^\circ$ at sunrise and $90^\circ$ when the sun is directly overhead. 
For 
each of the remaining 14 hours of the day, we build an ARMA$(p,q)$ 
model. 
For each hour, we verify the stationarity of the time 
series and test a number of ARMA($p,q$) models to find the best 
one. We use statistical tests on the residuals to validate the models. Finally, 
we use Monte Carlo sampling from the best ARMA model, for each hour, to create 
hourly scenarios. Below we provide more details.

\subsection{The ARMA model} \label{sec:arma}

We recall here to the general formulation of  ARMA models for modeling time 
series.  Given a time series $x_t$, we can model the level of its current 
observation depending on the level of its $p$ lagged observations, 
$x_{t-1},x_{t-2}, \ldots, x_{t-p}$, plus an additional white noise error term 
$\epsilon_t$. This model is known as an \textit{autoregressive} model of order 
${p}$, AR($p$): $x_t = \sum_{i=1}^p  \phi_i x_{t-i}   + \epsilon_t$.

Next, assume that the level of the current observation is affected not only 
by the current white noise error term, but also by the previous 
$q$ white noise errors. This model is known as a \textit{moving average} 
model of order $q$, MA($q$): $x_t = \epsilon_t + \sum_{i=1}^q  \theta_i \epsilon_{t-i}$.

An ARMA($p,q$) model combines both the AR($p$) and MA($q$) models as follows:
\begin{equation}\label{eq3}
x_t =  \sum_{i=1}^p  \phi_i x_{t-i}    + \epsilon_t + \sum_{i=1}^q  \theta_i \epsilon_{t-i}   
\end{equation}
Observe that the output variable depends linearly on the current and various 
past values (which is an advantage compared to other high fidelity 
forecasting models). An important question is how many 
representative lagged observations should be considered in order to have good 
fidelity while keeping the model as simple as possible. We discuss this in the 
proceeding sections.

\subsection{Stationarity} \label{sec:stationarity}
An ARMA model may be a suitable forecasting tool if a time-series is 
stationary. We test 
the hourly data for stationarity using the Augmented 
Dickey-Fuller (ADF) test~\cite{dickey1979distribution}. The ADF test has a null 
hypothesis that the series includes a unit root (or, is non-stationary). We 
reject the null hypothesis at a level 0.05 if the test-statistic exceeds its 
0.95 level quantile. For all the 14 hours of the day, the null hypothesis is 
rejected suggesting the series may be stationary, and hence an ARMA model may 
be suitable. If the series were not stationary, an ARIMA model may be suitable; 
see, e.g., \cite{contreras2003arima}.

\subsection{Selecting parameters of the ARMA model} 
Next, we estimate the parameters of the ARMA model, $p$, the order of the 
autoregressive part and, $q$, the order of the moving average part. For each 
hour, we construct 16 models with both $p$ and $q$ between one and four, and 
compute 
the log-likelihood objective function value. Next, for each hour, we calculate 
the Bayesian information criteria (BIC) for the 16 models using $p + q + 1$ 
parameters.  The BIC penalizes for models with more parameters, and the 
smallest value of the BIC gives the best model, for each 
hour. Table~\ref{tab:order} provides our estimated $p$ and $q$  values for 
the 14 hours of the day. We note that none of the hours have an order value 
exceeding two.

\begin{table}[!htb]
	\centering
	\caption{Estimated $p$ and $q$ values for ARMA($p,q$) models for 14 hours of 
		the day}
	\label{tab:order}
	\scalebox{0.8}{
		\begin{tabular}{c| cccccccccccccc}
			\hline 
			Hour & 6:00 & 7:00 & 8:00 & 9:00 & 10:00 & 11:00 & 12:00 \\ \hline
			p       & 1      & 1       & 1      & 1      & 2        & 1       & 1     \\
			q       & 1     & 1        & 1      & 1      & 1         & 1       & 2     \\ \hline
			
			& 13:00 & 14:00 & 15:00 & 16:00 & 17:00 & 18:00 & 19:00 \\ \hline
		    p      & 1     & 1     & 1  & 1     & 1     & 2     & 1   \\
			q	  & 1     & 1   & 1         & 1      & 1     & 1          & 2     \\ \hline

	\end{tabular}}
\end{table}

\subsection{Prediction}

\begin{figure}[!t]
\centering
\includegraphics[width=0.5\textwidth]{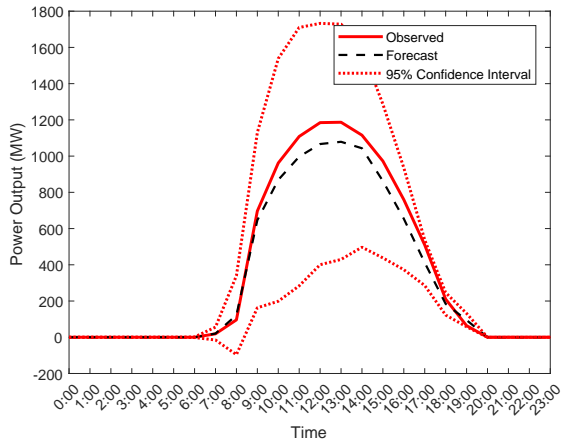}
\caption{Day-ahead actual and predicted values using ARMA models from Table~\ref{tab:order}}
\label{fig:prediction}
\end{figure}

Figure~\ref{fig:prediction} plots a day-ahead prediction using the above 
constructed ARMA models; i.e., one hour ahead predictions from the 14 ARMA 
models. A number of metrics are available to evaluate the prediction; see, 
e.g.,~\cite{coimbra2013overview}. We use a few of them here. The mean absolute 
error between the actual 
and the predicted series is 
39.6 MW, or  3.3\% of the maximum actual value. The root mean square 
error between the actual and the predicted series is 
61.0 MW, or  5.1\% of the maximum actual value. 

We further verify autocorrelation in the series, for each hour, using the 
Ljung-Box test~\cite{ljung1978measure} on the residuals for lags of 5, 10, 
and 15. The Ljung-Box  test has a 
null hypothesis that the residuals are uncorrelated up to a 
given lag. We reject the null hypothesis at a level 0.05 if its test-statistic 
exceeds its 0.95 level quantile. For all the 14 hours of the day, the null hypothesis is not
rejected suggesting a zero autocorrelation in the series, or the 
model choice may be appropriate.

\begin{figure}[!t]
	\centering
	\includegraphics[width=0.5\textwidth]{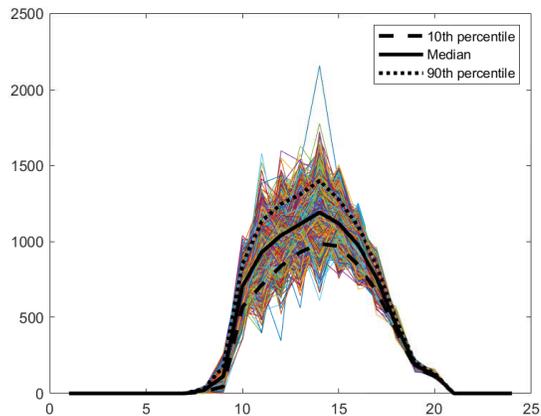}
	\caption{2000 hourly scenarios for solar power generated using the ARMA models 
		from Table~\ref{tab:order}. The dashed black line is the median hourly value, 
		and the solid black line is
		the 10 percentile solar power value.}
	\label{fig:sample}
\end{figure}

With increasing penetration of solar power in the electricity grid, a number of 
stochastic optimization models for bidding, storage, and generation have been 
developed; see, e.g.,~\cite{banos2011optimization}. 
Stochastic optimization models rely on the availability of a large number of 
scenarios.  We can use Monte Carlo sampling to generate hourly solar power 
scenarios. The output from an ARMA model is real valued, and hence can be 
negative. In our analysis, we truncate the negative powered outputs to 0. For 
the 14 hours of 
the day, this sampling resulted in 1.6\% of the outputs with estimated power 
output below -5MW. Figure~\ref{fig:sample} plots 2000 day-ahead scenarios 
as well as the median, 10 percentile, and 90 percentile values.

\section{Conclusions}
In this article, we present a simple step-by-step scheme for fitting an ARMA 
model to 
historical solar power data. The proposed model provides an easy-to-implement 
linear tool to forecast future hourly scenarios, or for embedding into 
decision-making models. 
We introduce statistical tests to check the applicability of various models, 
identify model parameters, and to finally forecast scenarios. If significant  statistical evidence is not present in support of a  test, the chosen model is not expected to perform well. This can lead to 
erroneous conclusions; see, 
e.g.,~\cite{kwiatkowski1992testing,phillips1988testing}. 
The proposed ARMA model performs better than a Smart-Persistence model (see Appendix). 
In summary, the methodology in this article can be directly applied to historical data,  both for a single PV source and a PV site,
to create future scenarios for use in stochastic or robust optimization models 
for power system operation and planning.

\section*{Appendix} \label{sec:appendix}

We compare the developed model against two other forecasting methods. First, we 
fit a single ARMA model, as opposed to hour-by-hour, on the entire time series. 
Second, we use the Smart-Persistence model of~\cite{lauret2016solar}, which 
assumes the solar power at hour $t$ is the mean of the previous $h$ hours. We 
use $h=2$ in here.
Table~\ref{tab:compare} compares the three models; the proposed model performs 
better in both of the chosen metrics. 

\begin{table}[h!]
\centering
\caption{Comparison of three forecasting models. MAE denotes the mean absolute error and RMSE denotes the root mean squared error.}
\label{tab:compare}
\begin{tabular}{c|ccc}
\toprule
\multicolumn{1}{l}{} & \multicolumn{3}{c}{Model}              \\ \hline
& Hourly ARMA & Single ARMA & Smart-Pers \\ \hline
MAE (MW)                 & 39.6        & 48.2        & 45.7   \\
RMSE (MW)                & 61          & 112.51      & 102.6   \\ \bottomrule
\end{tabular}
\end{table}

\section*{Acknowledgments} 
B.\ Singh thanks Jean-Paul Watson and Andrea Staid for helpful discussions and for sharing data. B.\ Singh's work was supported in part by Sandia's Laboratory Directed Research and Development (LDRD) program.
Sandia National Laboratories is a multimission laboratory managed and operated 
by National Technology and Engineering Solutions of Sandia, LLC., a wholly owned
subsidiary of Honeywell International, Inc., for the U.S. Department of 
Energy's 
National Nuclear Security Administration under contract DE-NA-0003525. This paper describes objective technical results and analysis. Any subjective 
views or opinions that might be expressed in the paper do not necessarily 
represent the views of the U.S. Department of Energy or the United States 
Government. 
D.\ Pozo's work was supported by Skoltech NGP Program (Skoltech-MIT joint 
project).

\bibliographystyle{plain}
\bibliography{mybibfile}

\begin{thebibliography}{10}

\bibitem{banos2011optimization}
Raul Banos, Francisco Manzano-Agugliaro, FG~Montoya, Consolacion Gil, Alfredo
  Alcayde, and Julio G{\'o}mez.
\newblock {O}ptimization methods applied to renewable and sustainable energy:
  {A} review.
\newblock {\em {R}enewable and {S}ustainable {E}nergy {R}eviews},
  15(4):1753--1766, 2011.

\bibitem{coimbra2013overview}
Carlos~FM Coimbra, Jan Kleissl, and Ricardo Marquez.
\newblock {O}verview of solar-forecasting methods and a metric for accuracy
  evaluation.
\newblock {\em {S}olar {E}nergy {F}orecasting and {R}esource {A}ssessment},
  pages 171--194, 2013.

\bibitem{contreras2003arima}
Javier Contreras, Rosario Espinola, Francisco~J Nogales, and Antonio~J Conejo.
\newblock {ARIMA} models to predict next-day electricity prices.
\newblock {\em {IEEE} transactions on power systems}, 18(3):1014--1020, 2003.

\bibitem{dickey1979distribution}
David~A Dickey and Wayne~A Fuller.
\newblock {D}istribution of the estimators for autoregressive time series with
  a unit root.
\newblock {\em {J}ournal of the {A}merican {S}tatistical {A}ssociation},
  74(366a):427--431, 1979.

\bibitem{dong2015novel}
Zibo Dong, Dazhi Yang, Thomas Reindl, and Wilfred~M Walsh.
\newblock {A} novel hybrid approach based on self-organizing maps, support
  vector regression and particle swarm optimization to forecast solar
  irradiance.
\newblock {\em {E}nergy}, 82:570--577, 2015.

\bibitem{duran2007short}
Mario~J Duran, Daniel Cros, and Jesus Riquelme.
\newblock {S}hort-term wind power forecast based on {ARX} models.
\newblock {\em {J}ournal of {E}nergy {E}ngineering}, 133(3):172--180, 2007.

\bibitem{golestaneh2016generation}
Faranak Golestaneh, Hoay~Beng Gooi, and Pierre Pinson.
\newblock {G}eneration and evaluation of space--time trajectories of
  photovoltaic power.
\newblock {\em {A}pplied {E}nergy}, 176:80--91, 2016.

\bibitem{graabak2016variability}
Ingeborg Graabak and Magnus Korp{\aa}s.
\newblock {V}ariability characteristics of {E}uropean wind and solar power
  resources---{A} review.
\newblock {\em {E}nergies}, 9(6):449, 2016.

\bibitem{huang2010comparative}
Yuehui Huang, Jing Lu, Chun Liu, Xiaoyan Xu, Weisheng Wang, and Xiaoxin Zhou.
\newblock {C}omparative study of power forecasting methods for {PV} stations.
\newblock In {\em Power System Technology (POWERCON), 2010 International
  Conference on}, pages 1--6. IEEE, 2010.

\bibitem{kleywegt2002sample}
Anton~J Kleywegt, Alexander Shapiro, and Tito Homem-de Mello.
\newblock The sample average approximation method for stochastic discrete
  optimization.
\newblock {\em SIAM Journal on Optimization}, 12(2):479--502, 2002.

\bibitem{kwiatkowski1992testing}
Denis Kwiatkowski, Peter~CB Phillips, Peter Schmidt, and Yongcheol Shin.
\newblock {T}esting the null hypothesis of stationarity against the alternative
  of a unit root: {H}ow sure are we that economic time series have a unit root?
\newblock {\em {J}ournal of {E}conometrics}, 54(1-3):159--178, 1992.

\bibitem{lauret2016solar}
Philippe Lauret, Elke Lorenz, and Mathieu David.
\newblock {S}olar forecasting in a challenging insular context.
\newblock {\em {A}tmosphere}, 7(2):18, 2016.

\bibitem{ljung1978measure}
Greta~M Ljung and George~EP Box.
\newblock {O}n a measure of lack of fit in time series models.
\newblock {\em {B}iometrika}, 65(2):297--303, 1978.

\bibitem{lorca2015adaptive}
Alvaro Lorca and Xu~Andy Sun.
\newblock Adaptive robust optimization with dynamic uncertainty sets for
  multi-period economic dispatch under significant wind.
\newblock {\em IEEE Transactions on Power Systems}, 30(4):1702--1713, 2015.

\bibitem{mora1998multiplicative}
LL~Mora-Lopez and M~Sidrach-de Cardona.
\newblock {M}ultiplicative {ARMA} models to generate hourly series of global
  irradiation.
\newblock {\em {S}olar {E}nergy}, 63(5):283--291, 1998.

\bibitem{phillips1988testing}
Peter~CB Phillips and Pierre Perron.
\newblock {T}esting for a unit root in time series regression.
\newblock {\em {B}iometrika}, 75(2):335--346, 1988.

\bibitem{technical}
{Solarfun Power}.
\newblock {SF} 160-24-{M}.
\newblock
  \url{https://www.energymatters.com.au/images/solarfun/SF160-24-M(190W).pdf},
  2009.
\newblock {Accessed: March 20, 2018}.

\bibitem{su2014stochastic}
Wencong Su, Jianhui Wang, and Jaehyung Roh.
\newblock {S}tochastic energy scheduling in microgrids with intermittent
  renewable energy resources.
\newblock {\em {IEEE} {T}ransactions on {S}mart {G}rid}, 5(4):1876--1883, 2014.

\bibitem{yang2018history}
Dazhi Yang, Jan Kleissl, Christian~A Gueymard, Hugo~TC Pedro, and Carlos~FM
  Coimbra.
\newblock {H}istory and trends in solar irradiance and {PV} power forecasting:
  {A} preliminary assessment and review using text mining.
\newblock {\em {S}olar {E}nergy}, 2018.

\end{thebibliography}

\end{document}